# Thermal wave crystals based on dual-phase-lag model

Zheng-Yang Li[a, b], Tian-Xue Ma[a, *], A-Li Chen[b], Yue-Sheng Wang[b, c, **], Chuanzeng Zhang[a]

[a] Department of Civil Engineering, University of Siegen, D-57076 Siegen, Germany

[b] Institute of Engineering Mechanics, Beijing Jiaotong University, Beijing 100044, China

[c] Department of Mechanics, School of Mechanical Engineering, Tianjin University, Tianjin 300350, China

**ABSTRACT**

Thermal wave crystals based on the dual-phase-lag model are investigated in this paper by both theoretical analysis and numerical simulation to control the non-Fourier heat conduction process. The transfer matrix method is used to calculate the complex dispersion curves. The temperature field is calculated by the finite difference time domain method. The results show that thermal band-gaps exist due to the Bragg-scattering. The key parameters characterizing the band-gaps are analyzed. The thermal wave impedance and mid-gap frequencies are introduced to predict band-gaps theoretically. Our results show that the larger the difference in the thermal wave impedances is, the wider of the thermal band-gaps will be. This study demonstrates a type of the thermal metamaterials which have potential innovative applications such as thermal imagining, thermal diodes and thermal waveguides for energy transmission.

**Keywords:** Non-Fourier heat conduction, Dual-phase-lag model, Thermal wave crystals, Transfer matrix method,

* Tianxue.Ma@uni-siegen.de
** yswang@tju.edu.cn

Band-gaps

## 1. Introduction

The manipulation of the heat transmission is fundamental to the development of many technological devices at micro- (e.g. electronic chips), meso- (e.g. engines) and macro-scales (e.g. rockets). The most traditional methods to control the heat conduction are based on defects and impurities. In recent years, the manipulation of electromagnetic waves by photonic crystals [1, 2] and acoustic/elastic waves by phononic crystals [3-5] have achieved great success. Correspondingly, by utilizing the coherent reflection of phonons, the heat flow can be transformed from a diffusive form to a wave phonon transport phenomenon. Based on this fact, Maldovan [6, 7] proposed the concept of 'thermal crystals' which aims to manage the heat flow. Recently, the heat guiding is realized by utilizing a gradient-like periodic structure hinged on the thermal crystals by Anufriev *et al* [8]. Despite the promising prospects of the thermal crystals, their limitations are also obvious, e.g., micro-scale, ultra-low temperature, etc.

The wave-like behavior of the thermal field exists not only at micro-scales. In 1958, Cattaneo [9] and Vernotte [10] individually presented a model, known as the Cattaneo-Vernotte (CV) heat conduction model, which introduces a time lag between the heat flux and the temperature gradient. The CV model dismisses the assumption of the instantaneous thermal propagation by the Fourier conduction law. However, the CV model still implicitly supposes an instant interaction between the temperature gradient and the energy transport. To overcome the drawbacks of the Fourier law and the CV model, Tzou [11-13] proposed the dual-phase-lag (DPL) model based on the time required by completing the physical interactions at micro- or nano-scales. By applying the first-order approximation, the heat conduction equation of the one-dimensional (1D) DPL model can be written as

$$q + \tau_q \frac{\partial q}{\partial t} = -\kappa \left( \frac{\partial T}{\partial x} + \tau_T \frac{\partial^2 T}{\partial x \partial t} \right), \qquad (1)$$

where $T$ is the temperature, $q = \partial T / \partial t$ is the heat flux, $x$ is the Cartesian coordinate, $t$ is the time, $\kappa$ is the thermal conductivity, $\tau_q$ is the phase-lag of the heat flux, and $\tau_T$ is the phase-lag of the temperature gradient.

The energy conservation equation is described by [14]

$$\frac{\partial q}{\partial x} = -\rho c_p \frac{\partial T}{\partial t} + Q, \qquad (2)$$

where $Q$ is the internal energy generation rate, $\rho$ is the mass density and $c_p$ is the specific heat. Combining Eq. (1) with Eq. (2) yields [14]

$$\frac{1}{\tau_q}\frac{\partial T}{\partial t} + \frac{\partial^2 T}{\partial t^2} = \frac{\kappa}{\rho c_p \tau_q}\frac{\partial^2 T}{\partial x^2} + \frac{\kappa \tau_T}{\rho c_p \tau_q}\frac{\partial^3 T}{\partial x^2 \partial t} + \frac{1}{\kappa}\left(Q + \tau_q \frac{\partial Q}{\partial t}\right). \qquad (3)$$

Eq. (3) is a hyperbolic equation which assumes that the heat in the DPL model propagates with the same speed as in the CV model, i.e., $C_{\text{DPL}} = C_{\text{CV}} = \sqrt{\kappa(\rho c_p \tau_q)^{-1}}$.

Some other models were proposed to describe the heat conduction process more specifically, e.g., the thermomass model [15, 16], the extended irreversible thermodynamics model [17, 18], and the alternative approaches to the analysis of the diffusion equation [19, 20]. Due to the peculiarity of the resulting hyperbolic wave equation, the wave-like behavior of the thermal transfer has been investigated in the past several decades, such as the heat conduction in the functional gradient materials by Rahideh *et al* [21]. Reviews on the DPL theory were given by Ozisik *et al.* [22], Xu *et al.* [23] and Fan *et al.* [24] among others. Zhang *et al.* [25] presented two exact explicit solutions for the three-dimensional DPL heat conduction equation. Chou and Yang [26, 27] examined the DPL thermal behavior in 1D and two-dimensional (2D) single-/multi-layered structures. Zhang *et al.* [28] analyzed the pulsed laser effect on skin by the DPL model utilizing a three-level finite difference method.

In this article, we discuss the non-Fourier heat conduction in a 1D periodic structure based on the DPL model and Bloch-theorem [29]. Our previous research [30] has demonstrated that the thermal band-gaps, which block the heat transfer, exist in the thermal wave crystals analog to the photonic crystals and the phononic crystals. This research work aims to examine whether thermal band-gaps can appear in the thermal wave crystals described by the DPL model. This research work provides a more generalized way to manipulate the heat transfer or thermal waves, analog to the electromagnetic wave manipulation by the photonic crystals [31] and the acoustic/elastic wave manipulation by the phononic crystals [32], respectively.

## 2. Problem formulation

A 1D thermal wave crystal consisting of bilayer unit-cells is considered, as shown in Fig. 1. Each of the unit-cells contains sub-cells A and B with the thicknesses $l_A$ and $l_B$, respectively, and the total unit-cell's thickness is $l = l_A + l_B$. The material properties ($\kappa$, $\rho$, $c_p$, $\tau_q$, $\tau_T$, $C_{DPL}$) of the two sub-layers of the unit-cells are marked by the subscripts A and B, and the coordinate system is defined in Fig. 1. By considering an arbitrary unit-cell such as the $j$th unit-cell, the left and right boundary coordinates of this unit-cell are $x_L^j = jl$ and $x_R^j = (j+1)l$, and the coordinate of the interface between A and B is $x_{AB}^j = jl + l_A$.

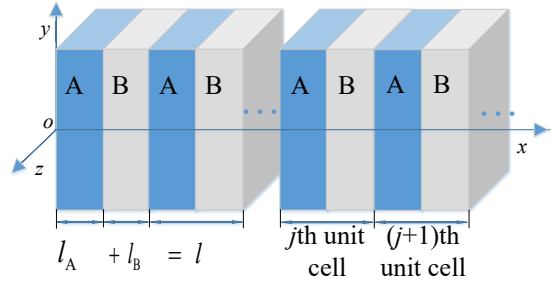

Fig. 1. Illustration of the 1D thermal wave crystal.

For a time-harmonic thermal wave with angular frequency $\omega$ propagating in the 1D thermal wave crystal without internal heat source or loss (i.e. $Q = 0$), the temperature and heat flux can be denoted as $\{T(x,t), q(x,t)\} = \{\hat{T}(x), \hat{q}(x)\}e^{-i\omega t}$. Then, the heat conduction Eq. (3) can be rewritten as the following differential equation for the temperature amplitude $\hat{T}(x)$:

$$\hat{T}''(x) + \frac{\omega^2 + i\omega/\tau_q}{C_{DPL}^2(1 - i\omega\tau_T)}\hat{T}(x) = 0, \tag{4}$$

where $i = \sqrt{-1}$. The solution of Eq. (4) for $\hat{T}(x)$ can be written as

$$\hat{T}(x) = A_1 e^{i\gamma x} + A_2 e^{-i\gamma x}, \tag{5}$$

where $A_1$ and $A_2$ are the unknown coefficients, and

$$\gamma = \sqrt{\frac{\omega^2 + i\omega/\tau_q}{C_{DPL}^2(1-i\omega\tau_T)}}. \tag{6}$$

The real and imaginary parts of Eq. (6) describe the propagation and attenuation of the thermal wave, respectively. The heat flux $\hat{q}(x)$ can be obtained from Eqs. (1) and (5) as

$$\hat{q}(x) = -A_1 \frac{i\kappa(i+\omega\tau_T)\gamma}{1-i\omega\tau_q} e^{i\gamma x} + A_2 \frac{i\kappa(i+\omega\tau_T)\gamma}{1-i\omega\tau_q} e^{-i\gamma x}. \tag{7}$$

For the sake of brevity, the state vector is introduced as

$$\mathbf{S}(x) = \{\hat{T}(x), \hat{q}(x)\}^T = \begin{bmatrix} e^{i\gamma x} & e^{-i\gamma x} \\ \dfrac{i\kappa(i+\omega\tau_T)\gamma}{1-i\omega\tau_q}e^{i\gamma x} & \dfrac{i\kappa(i+\omega\tau_T)\gamma}{1-i\omega\tau_q}e^{-i\gamma x} \end{bmatrix} \begin{bmatrix} A_1 \\ A_2 \end{bmatrix} = \mathbf{M}(x)\{A_1, A_2\}^T, \tag{8}$$

where the superscript T represents the transpose of a vector. The state vector is denoted as $\mathbf{S}_A^j(x) = \mathbf{M}_A^j(x)\{A_1, A_2\}^T$ (with $x_L^j < x < x_{AB}^j$) for the layer A of the $j$th unit-cell, and $\mathbf{S}_B^j(x) = \mathbf{M}_B^j(x)\{B_1, B_2\}^T$ (with $x_{AB}^j < x < x_R^j$) for the layer B. The matrices $\mathbf{M}_A^j(x)$ and $\mathbf{M}_B^j(x)$ in Eq. (8) are obtained from the corresponding material parameters for the sub-layers A and B, respectively.

Next, the transfer matrix (TM) method [33] is used to calculate the dispersion curves. By defining the state vectors at the boundaries of the layers A and B as $\mathbf{S}_{AL}^j = \mathbf{S}_A^j(x_L^j)$, $\mathbf{S}_{AR}^j = \mathbf{S}_A^j(x_{AB}^j)$, $\mathbf{S}_{BL}^j = \mathbf{S}_B^j(x_{AB}^j)$ and $\mathbf{S}_{BR}^j = \mathbf{S}_B^j(x_R^j)$, then we obtain from Eq. (8): $\mathbf{S}_{AL}^j = \mathbf{M}_A^j(x_L^j)\{A_1, A_2\}^T$, $\mathbf{S}_{AR}^j = \mathbf{M}_A^j(x_{AB}^j)\{A_1, A_2\}^T$, $\mathbf{S}_{BL}^j = \mathbf{M}_B^j(x_{AB}^j)\{B_1, B_2\}^T$, and $\mathbf{S}_{BR}^j = \mathbf{M}_B^j(x_R^j)\{B_1, B_2\}^T$. Eliminating $\{A_1, A_2\}^T$ and $\{B_1, B_2\}^T$ yields

$$\mathbf{S}_{AR}^j = \mathbf{M}_{AR}^j(\mathbf{M}_{AL}^j)^{-1}\mathbf{S}_{AL}^j, \quad \mathbf{S}_{BR}^j = \mathbf{M}_{BR}^j(\mathbf{M}_{BL}^j)^{-1}\mathbf{S}_{BL}^j, \tag{9}$$

where $\mathbf{M}_{AL}^j = \mathbf{M}_A^j(x_L^j)$, $\mathbf{M}_{AR}^j = \mathbf{M}_A^j(x_{AB}^j)$, $\mathbf{M}_{BL}^j = \mathbf{M}_B^j(x_{AB}^j)$ and $\mathbf{M}_{BR}^j = \mathbf{M}_B^j(x_R^j)$.

The temperature and heat flux are continuous at the interface between two adjacent sub-layers, i.e.,

$$\mathbf{S}_{AR}^j = \mathbf{S}_{BL}^j, \quad \mathbf{S}_{BR}^j = \mathbf{S}_{AL}^{j+1}. \tag{10}$$

Substituting Eq. (9) into Eq. (10) and eliminating $\mathbf{S}_{BL}^j$, we obtain

$$\mathbf{S}_{AL}^{j+1} = \mathbf{M}_{BR}^j(\mathbf{M}_{BL}^j)^{-1}\mathbf{M}_{AR}^j(\mathbf{M}_{AL}^j)^{-1}\mathbf{S}_{AL}^j = \mathbf{M}_{\text{Transfer}}^j \mathbf{S}_{AL}^j, \tag{11}$$

which connects the state vectors of the $j$th and $(j+1)$th unit-cells. The matrix $\mathbf{M}_{\text{Transfer}}^{j}$ is the transfer matrix between two consecutive unit-cells, which is the same for any value of $j$ and therefore denoted as $\mathbf{M}_{\text{Transfer}}$.

In our previous work [30], we have shown that the Bloch-theorem [34] can be applied to describe the non-Fourier heat conduction process in periodic media. Accordingly, the following periodic condition for the right and left boundaries of the $j$th unit-cell of the 1D thermal wave crystal can be applied:

$$\mathbf{S}_{\text{AL}}^{j+1} = e^{ikl}\mathbf{S}_{\text{AL}}^{j}, \qquad (12)$$

where $k$ is the complex Bloch wave number, which can be written as $k=k_{\text{R}}+ik_{\text{I}}$, where $k_{\text{R}}$ and $k_{\text{I}}$ are the real and imaginary parts of $k$, respectively. Substituting Eq. (12) into Eq. (11) yields the following eigenvalue equation:

$$\mathbf{M}_{\text{Transfer}}^{j}\mathbf{S}_{\text{AL}}^{j} = e^{ikl}\mathbf{S}_{\text{AL}}^{j}. \qquad (13)$$

Substituting Eqs. (5)-(11) into Eq. (13), one can obtain the following characteristic equation [33]:

$$\cosh(ikl) = \cosh(i\gamma_{\text{A}}l_{\text{A}})\cosh(i\gamma_{\text{B}}l_{\text{B}}) + \frac{1}{2}\left(\frac{\eta_{\text{A}}\gamma_{\text{A}}}{\eta_{\text{B}}\gamma_{\text{B}}} + \frac{\eta_{\text{B}}\gamma_{\text{B}}}{\eta_{\text{A}}\gamma_{\text{A}}}\right)\sinh(i\gamma_{\text{A}}l_{\text{A}})\sinh(i\gamma_{\text{B}}l_{\text{B}}), \qquad (14)$$

where $\eta_{\text{A}} = \kappa_{\text{A}}(i+\omega\tau_{\text{TA}})(1-i\omega\tau_{q\text{A}})^{-1}$ and $\eta_{\text{B}} = \kappa_{\text{B}}(i+\omega\tau_{\text{TB}})(1-i\omega\tau_{q\text{B}})^{-1}$. By solving Eq. (14), the complex dispersion relations between the angular frequency $\omega$ and the complex wave number $k$ can be obtained.

The dispersion curve, also known as the band structure, is generally represented within the first Brillouin-zone of a periodic system and used to characterize the inherent features of the waves [35]. A 'band-gap' can be defined as the frequency range in which the imaginary part of $k$ is larger than the average value of the imaginary parts of the bulk materials A and B. In a band-gap, all Bloch waves are evanescent, which decay faster than the average rate of the bulk materials A and B [32, 34, 35]. Otherwise, it is called a 'pass-band'. In the next section, numerical results for the thermal waves in a 1D thermal wave crystal will be presented and discussed in details.

## 3. Numerical results and discussion

For the numerical calculations, we consider a thermal wave crystal with $l_{\text{A}} = l_{\text{B}} = 0.01$ mm. The Stratum-like and Dermis-like materials are chosen, and the material parameters are listed in Table 1 [36, 37].

Table 1 Material constants [36, 37]

| Component materials | Stratum-like (Layer A) | Dermis-like (Layer B) |
|---|---|---|
| Thermal conductivity (W m$^{-1}$ K$^{-1}$) | $\kappa_A = 0.235$ | $\kappa_B = 0.445$ |
| Specific heat (J kg$^{-1}$ K$^{-1}$) | $c_{pA} = 3600$ | $c_{pB} = 3300$ |
| Density (kg m$^{-3}$) | $\rho_A = 1500$ | $\rho_B = 1116$ |
| Phase-lag of the heat flux vector (s) | $\tau_{qA} = 1$ | $\tau_{qB} = 20$ |
| Phase-lag of the temperature gradient (s) | $\tau_{TB} = 0.0002$ | $\tau_{TB} = 0.004$ |

The complex dispersion curves are illustrated in Figs. 2(a) and 2(b), which correspond, respectively, to the real and imaginary parts of the complex wave number. It needs to be clarified that the real part in Fig. 2(a) never turns into zero or 1.0. In Fig. 2(b), the imaginary part of the normalized wave number almost equals to the average values of A and B in the frequency ranges of 0-1.8 Hz, 3.5-4.7 Hz and 6.7-7.9 Hz. This implies that the periodicity is not the main reason causing the attenuation, and therefore these frequency ranges are called pass-bands. However, in the frequency ranges of 1.8-3.4 Hz, 4.8-6.7 Hz and 7.9-9.3 Hz (the gray areas in Fig. 2(b)), the imaginary part is larger than the average of the imaginary parts of the bulk materials A and B. It means that the periodicity of the structure is the primary cause of the attenuation. These frequency ranges are called band-gaps, whose boundaries can be determined by the extremum value of the derivative of the imaginary part as shown in Fig. 2(c). As the frequency gets higher, e.g., over 9 Hz, the imaginary part of the normalized wave number converges to that of the average value of A and B, implying that the effect of the periodicity is submerged by the damping effect of the DPL. That turns all the frequency ranges over 9 Hz into pass-bands.

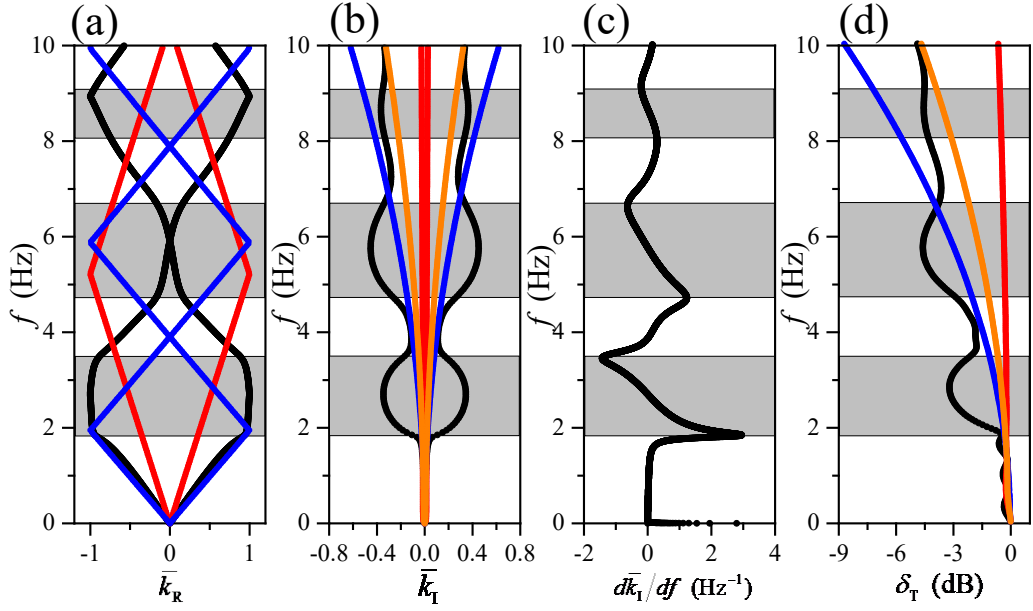

Fig. 2. Complex dispersion curves of the 1D thermal wave crystal (the black solid curve), bulk material A (the red solid curve), bulk material B (the blue solid curve) and average of A and B (the orange solid curve) for the real (a) and imaginary (b) parts of the normalized wave number $\bar{k} = \bar{k}_R + i\bar{k}_I = kl/\pi$. The derivative of $\bar{k}_I$ is shown in (c), and its maxima and minima determine the lower and upper edges of the band-gaps. (d) The temperature response $\delta_T$ is calculated by the FDTD method for the thermal wave crystal (the black solid curve), homogeneous bulk material A (the red solid curve), bulk material B (the blue solid curve) and average of A and B (the orange solid curve).

Then the finite difference time domain (FDTD) method [38, 39] is applied to verify and explain the results calculated by the TM method. A finite thermal wave crystal with 4 unit-cells and connected with a homogeneous medium A of the length $l$ is calculated. The thermal loading ($T(0,t) = T_0 \sin(2\pi f t) + 35°C$) is applied as input and the details of the FDTD method can be found in the appendix. The temperature response as output is defined as the ratio between the amplitudes of the input and output temperatures given by $\delta_T = \ln[|T_{x=0}|/|T_{x=4l}|]$, which is plotted in Fig. 2(d) (see the black solid curve). The results for homogeneous bulk material A (the red solid curve) and B (the blue solid curve) with the same thickness and their average (the orange solid curve) are also shown in Fig. 2(d). It can be clearly seen that the temperature attenuates rapidly in the band-gaps shown in Figs. 2(a) or 2(b), and that a larger value of the imaginary part of the wave number corresponds to a higher attenuation of the temperature. In a word, the thermal wave crystal shows more pronounced heat attenuation than that of the homogeneous material in the band-gaps due to the thermal wave effect and the periodicity.

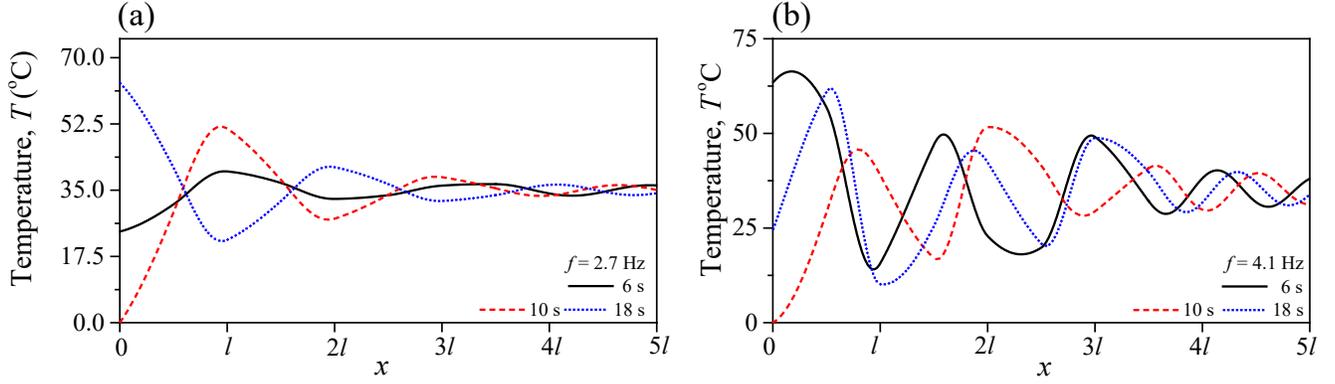

Fig. 3. The temperature distributions calculated by the FDTD method: (a) $f$ = 2.7 Hz and (b) $f$ = 4.1 Hz; $f$ is the frequency of the thermal loading.

Figure 3 illustrates the temperature distributions with the loading frequency being 2.7 Hz inside the first band-gap (Fig. 2) and 4.1 Hz outside the band-gap (Fig. 2). In Fig. 3, the amplitude values of the temperature attenuate strongly within the band-gap (Fig. 3(a)) while it can propagate through the structure outside the band-gap (Fig. 3(b)). This abnormal behavior cannot be found in the bulk materials A and B, and cannot be predicted by the average value of A and B, which means that the periodicity of the structure is the main reason for this abnormal behavior. The wave-length of the first band-gap in Fig. 3(a) is about 2 times of the periodicity of the structure, indicating that the generation of the first band-gap stems from the Bragg-scattering of the thermal wave. As the non-Fourier heat transfers through the periodic structure, heat reflections and transmissions happen at the interfaces. In specific frequency ranges, the reflected thermal waves and the incident thermal waves interact destructively, which then prevents the propagation of the thermal waves. For a Bragg-scattering type band-gap, the mid-gap wave-length is about $m/2$ times the length of the unit-cell, where $m$ is the order number of the band-gap [40]. Correspondingly, the mid-gap frequency in the 1D thermal wave crystal is

$$f_{\text{Center}}^{(m)} = \frac{m}{2(l_A/C_{\text{DPLA}} + l_B/C_{\text{DPLB}})}, \qquad (15)$$

which yields $f_{\text{Center}}^{(1)}$ = 2.8 Hz for $m=1$, $f_{\text{Center}}^{(2)}$ = 5.6 Hz for $m=2$, and $f_{\text{Center}}^{(3)}$ = 8.4 Hz for $m=3$. These values are close to the mid-gap frequencies in Fig. 2 (2.7 Hz, 5.7 Hz and 8.6 Hz for the first, second and third band-gaps, respectively). That means that all band-gaps are generated by the Bragg-scattering mechanism.

## 4. Parameter analysis

To characterize the effect of the periodicity on the non-Fourier heat conduction progress, we introduce the thermal wave impedance ratio [41] as

$$\varphi = \frac{w_A}{w_B} = \frac{c_{pA}\rho_A C_{DPLA}}{c_{pB}\rho_B C_{DPLB}} = \sqrt{\frac{\bar{\tau}_{qn}}{\bar{\kappa}_n \bar{c}_{\rho n}}}, \quad (16)$$

where $w_A$ and $w_B$ are the thermal wave impedances for the layers A and B, respectively, $\bar{\kappa}_n = \kappa_B \kappa_A^{-1}$ is the thermal conductivity ratio, $\bar{c}_{\rho n} = \rho_B c_{pB} (\rho_A c_{pA})^{-1}$ is the volumetric thermal capacity ratio, and $\bar{\tau}_{qn} = \tau_{qB} \tau_{qA}^{-1}$ is the thermal relaxation time ratio. Eq. (16) can be used to characterize whether the thermal wave reflection at the interface is strong or weak. When $\varphi = 1$, the wave reflection is totally eliminated, which means that the band-gap no longer exists. Correspondingly, the non-dimensional mid-gap frequency $\bar{f} = f_{Center}^{(m)} l C_{DPLA}^{-1}$ is derived as

$$\bar{f} = \frac{m}{2\bar{l}\left[n_A + (1-n_A)\sqrt{\bar{c}_{\rho n}\bar{\tau}_{qn}/\bar{\kappa}_n}\right]}, \quad (17)$$

where $\bar{l} = l(C_{DPLA}\tau_{qA})^{-1}$ is the non-dimensional length, $n_A = l_A l^{-1}$ and $n_B = l_B l^{-1}$ are the filling ratios of the layers A and B ($n_A + n_B = 1$), respectively.

From Eq. (17), the key non-dimensional parameters [41] characterizing the band-gaps are $\bar{l}$, $n_A$ (or $n_B$), $\bar{\kappa}_n$, $\bar{c}_{\rho n}$ and $\bar{\tau}_{qn}$. For the previously considered example, the corresponding parameters are 0.0959, 0.5, 1.894, 0.682, 20, respectively. Next, the influences of these parameters on the band-gaps are analyzed by altering the value of one parameter while keeping the others fixed. The results are illustrated in the following figures with the color representing the value of the imaginary part of $\bar{k}$, where the bright colors mean larger values. The band-gaps and non-dimensional mid-gap frequencies are also depicted with the black dashed lines in the figures.

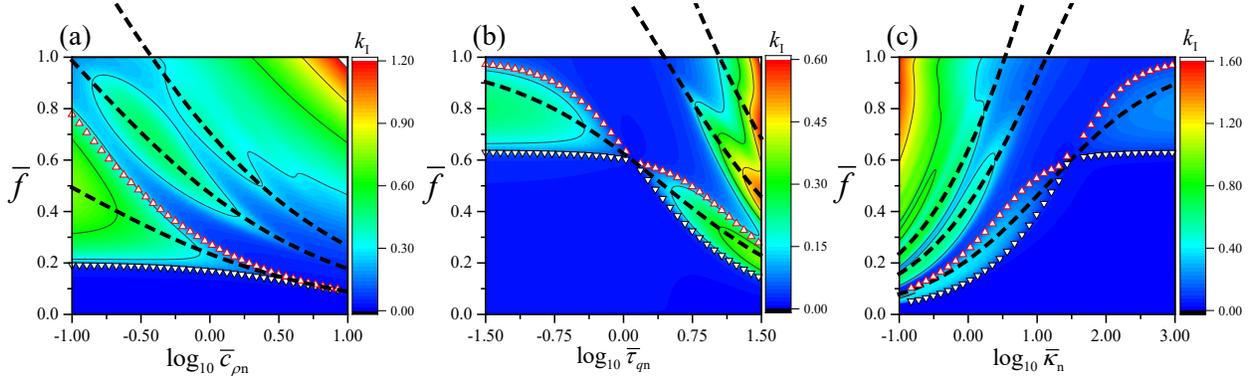

Fig. 4. Influences of the volumetric thermal capacity ratio $\bar{c}_{\rho n}$ (a), the thermal relaxation time ratio $\bar{\tau}_{qn}$ (b), and the thermal conductivity ratio $\bar{\kappa}_n$ (c) on the imaginary part of the wave number $\bar{k}$. The color represents the value of $\bar{k}_I$, the up triangle/down triangles show the upper/lower edges of the first band-gap, and the dashed lines are the mid-gap frequencies calculated by Eq. (17).

Figure 4 shows the effects of the thermal capacity ratio $\bar{c}_{\rho n}$, the relaxation time ratio $\bar{\tau}_{qn}$ and the thermal conductivity ratio $\bar{\kappa}_n$ on the imaginary part of the wave number. The mid-gap frequencies decrease as $\bar{c}_{\rho n}$ (Fig. 4(a)) or $\bar{\tau}_{qn}$ (Fig. 4(b)) increases, while increase as $\bar{\kappa}_n$ (Fig. 4(c)) increases. This phenomenon can be explained by considering the occurrence of $\bar{c}_{\rho n}\bar{\tau}_{qn}\bar{\kappa}_n^{-1}$ in Eq. (17).

The widening and narrowing of the band-gaps can be understood by considering Eq. (16). The band-gap will be wider when $\varphi$ is larger or smaller than 1, and will disappear when $\varphi = 1$. Suppose $\bar{\tau}_{qn}$ and $\bar{\kappa}_n$ are constants and $\varphi = 1$, we can obtain $\log_{10} \bar{c}_{\rho n} = 1.024$ from Eq. (16). The same process can be followed to obtain $\log_{10} \bar{\tau}_{qn} = 0.111$ and $\log_{10} \bar{\kappa}_n = 1.467$. These results show good agreements with Fig. 4.

Figure 5(a) shows the influence of the non-dimensional length $\bar{l}$ on the imaginary part of the wave number. It is rather obvious that the band-gaps do not vary with $\bar{l}$ in the range $-1.25 < \log_{10} \bar{l} < 0.5$. This implies that the scaling law, i.e., uniformly increasing or reducing the geometrical sizes of the structures by a factor $\varepsilon$ results in the frequency spectrum being scaled by $1/\varepsilon$, holds in this case. This result again proves that the band-gaps are generated by the Bragg-scattering. This is also evidenced by Eq. (17) which is independent of $\bar{l}$. When $\log_{10} \bar{l} > 1$ or $\log_{10} \bar{l} < -1.25$, all of the disappeared band-gaps can be described by the 'critical damping' or 'over damping' as in the theory of mechanical vibration [14]. The critical damping and over damping makes the non-Fourier heat conduction process more like a diffusion process, which means that the attenuation increases remarkably as the frequency increases [42]. In a word, the non-dimensional length $\bar{l}$ has a significant effect on the band-gap properties or the thermal wave propagation when the thermal wave crystal is over damped. While the non-dimensional length $\bar{l}$ has no effect on the band-gap when the thermal wave crystal is under damped.

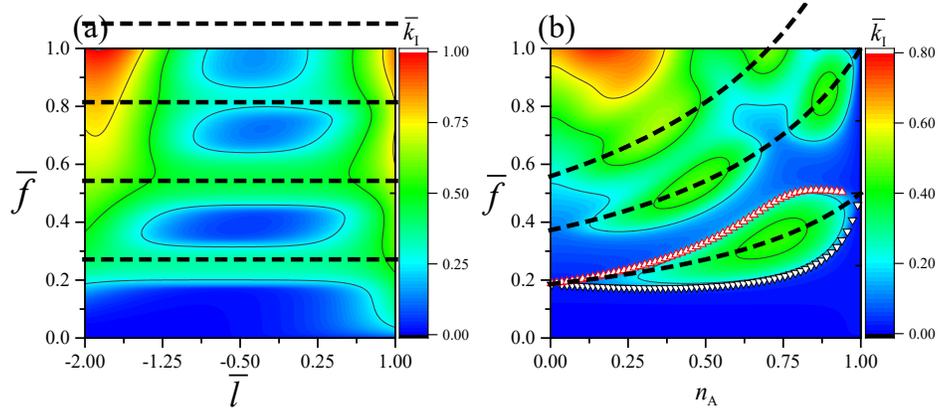

Fig. 5. Effects of the non-dimensional length $\bar{l}$ (a) and the filling ratio $n_A$ (b) on the imaginary part of the wave number $\bar{k}$. The color represents the value of the imaginary part of $\bar{k}$, the up triangle/down triangles show the upper/lower edges of the first band-gap, the dashed lines are the mid-gap frequencies calculated by Eq. (17).

The effect of the filling ratio $n_A$ of the layer A on the band-gaps is shown in Fig. 5(b). The band-gaps become wider first and then narrower as $n_A$ increases. The band-gaps disappear when $n_A = 0$ or $1$, which means that the periodicity is the key factor to generate a band-gap. The mid-gap frequencies increase monotonically with the increase of $n_A$, which can be explained by $\sqrt{\bar{c}_{\rho n} \bar{\tau}_{qn} \bar{\kappa}_n^{-1}} > 1$ in Eq. (17). Note that the lower and upper boundaries of the first band-gap will be converged to the frequencies calculated by Eq. (17) as the filling ratio approaches 0 or 1.

Although Eqs. (16) and (17) show that the phase-lag of the temperature gradient has no effect on the generation of the band-gaps, we should still analyze its effect on the dispersion curves. Here we introduce the ratio between two phase lags as $\beta_A = \tau_{TA}/\tau_{qA}$ and $\beta_B = \tau_{TB}/\tau_{qB}$ for materials A and B, respectively. Figure 6 shows the influences of the phase-lag ratio $\beta_{A/B}$ on the imaginary part of the wave number.

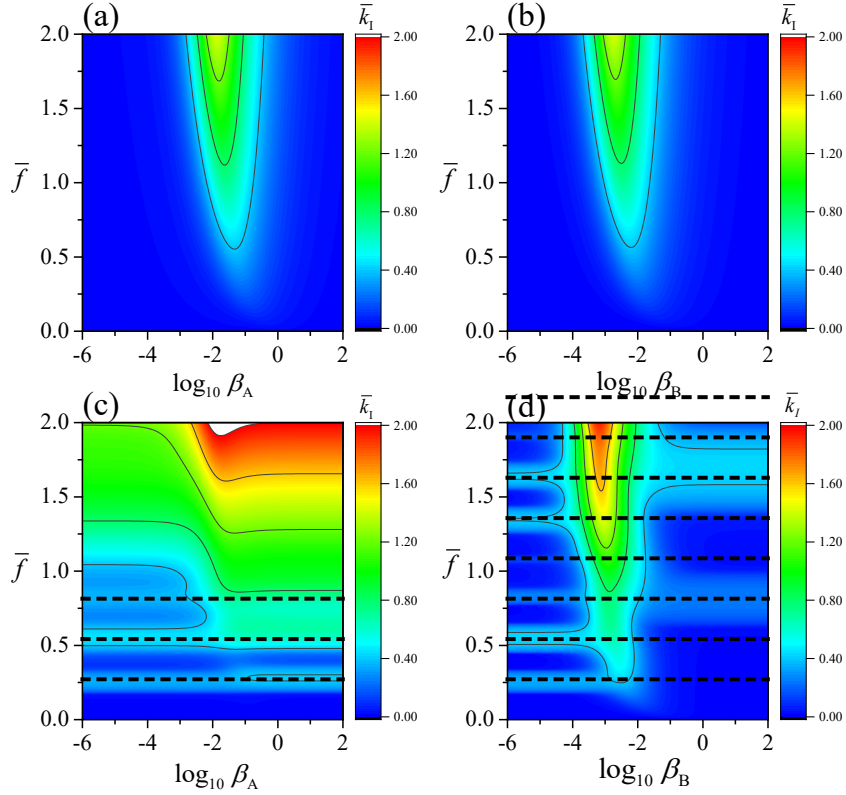

Fig. 6. Effects of the phase-lag ratio $\beta$ in the bulk material A (a) and the bulk material B (b) on the imaginary part of the wave number $\bar{k}$. Effects of the phase-lag ratio $\beta$ of the material A $\beta_A$ (c), and the material B $\beta_B$ (d) in the thermal wave crystal on the imaginary part of the wave number $\bar{k}$. The color represents the value of the imaginary part of $\bar{k}$ and the dashed lines are the mid-gap frequencies calculated by Eq. (17).

The increase of the imaginary part of the non-dimensional wave number $\bar{k}_I$ is influenced by $\beta_{A/B}$ as shown in Fig. 6, and $\bar{k}_I$ has extreme values at certain values of $\beta_{A/B}$, such as $\log_{10}\beta_A = 1.8$ in Fig. 6(a), $\log_{10}\beta_B = 2.9$ in Fig. 6(b), $\log_{10}\beta_A = 1.9$ in Fig. 6(c) and $\log_{10}\beta_B = 3$ in Fig. 6(d). For a given non-dimensional frequency, $\bar{k}_I$ has the same maximum and minimum for the bulk materials A (see Fig. 6(a)) and B (see Fig. 6(b)). The phase-lag ratio $\beta_{A/B}$ may demolish some band-gaps (see Fig. 6(d)) but does not change the frequencies of the band-gaps (see Figs. 6(c) and 6(d)). This phenomenon implies that $\beta$ acts like the viscosity for elastic waves in the Kelvin-Voigt model [43].

## 5. Concluding remarks

In this article, 1D thermal wave crystals described by the DPL model to control the non-Fourier heat conduction process are investigated. The complex dispersion curves are calculated by the TM method with complex Bloch wave numbers. The transient temperature field in the thermal wave crystals is calculated by the FDTD method. The results show that the Bragg-scattering band-gaps do exist in the non-Fourier heat transfer process described by the DPL model. These band-gaps are predicted analytically by the thermal wave impedance ratio and the mid-gap frequencies.

The results indicate that the larger the difference between the sub-layers' thermal wave impedances is, the wider the band-gaps will be. The mid-gap frequencies are determined by the filling ratio $n$, the thermal conductivity ratio $\bar{\kappa}_n$, the volumetric thermal capacity ratio $\bar{c}_{\rho n}$ and the thermal relaxation time ratio $\bar{\tau}_{qn}$. The non-dimensional length $\bar{l}$ and the ratio between the two phase-lags $\beta$ determine the under damping, critical damping and over damping of the system. The use of the non-dimensional length also ensures that the obtained results should follow the scaling law.

This work demonstrates that the thermal manipulation can be realized by the thermal wave crystals, like the light or acoustic/elastic wave manipulation in photonic or phononic crystals. This research indicates new ways to design thermal metamaterials for thermal energy transmission, conservation and management.

**Acknowledgements**

The work was supported by the German Research Foundation (DFG, Project-No. ZH 15/30-1, ZH 15/27-1) and the Joint Sino-German Research Project (Grant No. GZ 1355).

**Appendix**

The FDTD method with absorbing boundary conditions is used to calculate the temperature response in a thermal wave crystal with finite unit-cells. The finite 1D layered system with 4 periodic unit-cells connected to a homogeneous layer of the component material A is considered, see Fig. A1. In all numerical calculations we select $b = 4l$ and $d = 5l$, and a specific temperature loading is applied at the boundary of $x = 0$ to generate a time-harmonic temperature or heat flux disturbance. The particular forms of the initial and boundary conditions are written as

$$T(x,t) = 35\,°C, \quad q(x,t) = 0, \text{ at } t = 0. \tag{A1}$$

$$T(x,t) = T_0\left[1 - \cos(2\pi f t)\right]e^{\frac{-4\pi(t-0.8n)^2}{s^2}} + 35\,°C, \quad x = 0, \tag{A2}$$

$$T(x,t) = 35\,°C, \text{ at } x = d, \tag{A3}$$

$$T_A(x,t) = T_B(x,t), \quad q_A(x,t) = q_B(x,t), \text{ at each interface between two sub-layers,} \tag{A4}$$

where $T_0$ is the amplitude of the temperature change, $n$ and other parameters except $f$ are properly selected in particular examples. In calculating Fig. 2(d) and Fig. 3, we choose $s = 8$.

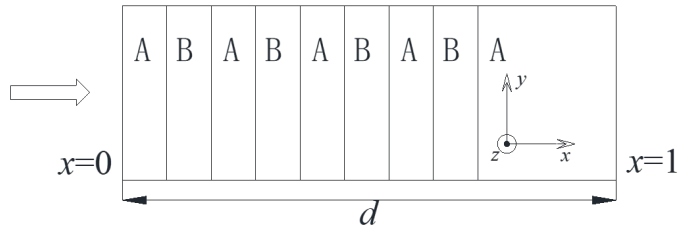

Fig. A1. Schematic diagram of the 1D heat conduction problem

To compare the results described by the dispersion curves for an infinite system, we adopt an absorbing boundary condition [44] at the right end of the structure to eliminate the reflected waves.

By rewriting Eq. (4) as

$$C_{DPL}^2 \frac{\partial^2 \hat{T}}{\partial x^2} + \frac{\omega^2 + i\omega\tau_q^{-1}}{1 - i\omega\tau_T} \hat{T} = 0, \tag{A5}$$

or

$$\left[ C_{DPL} \frac{\partial}{\partial x} - i\sqrt{\frac{\omega^2 + i\omega\tau_q^{-1}}{1 - i\omega\tau_T}} \right] \cdot \left[ C_{DPL} \frac{\partial}{\partial x} + i\sqrt{\frac{\omega^2 + i\omega\tau_q^{-1}}{1 - i\omega\tau_T}} \right] \hat{T} = 0, \tag{A6}$$

we can find that either of the following two equations should be satisfied:

$$\left( C_{DPL} \frac{\partial}{\partial x} + i\sqrt{\frac{\omega^2 + i\omega\tau_q^{-1}}{1 - i\omega\tau_T}} \right) A_1 e^{i\gamma x - i\omega t} = 0, \tag{A7}$$

$$\left( C_{DPL} \frac{\partial}{\partial x} - i\sqrt{\frac{\omega^2 + i\omega\tau_q^{-1}}{1 - i\omega\tau_T}} \right) A_2 e^{-i\gamma x - i\omega t} = 0. \tag{A8}$$

These two equations describe two thermal waves propagating in opposite directions. We consider the one propagating along the positive x-direction, i.e., Eq. (A7). When $\omega\tau_q^{-1}$ and $\omega\tau_T$ are very small, the imaginary parts $i\omega\tau_q^{-1}$ and $i\omega\tau_T$ in Eq. (A7) can be neglected. Then it is easy to get the time-domain absorbing boundary condition at $x = d$ as [30]

$$\left( C_{DPL} \frac{\partial}{\partial x} + \frac{\partial}{\partial t} \right) T = 0. \tag{A9}$$

Although this absorbing boundary condition is obtained by assuming $\tau_q \gg 1/\omega$ and $\tau_T \ll \omega$, it can yield acceptable results for $\tau_q \sim 1/\omega$ and $\tau_T \sim \omega$. In fact, when $\tau_q$ is very small or $\tau_T$ is very large, the attenuation of the thermal wave is significant and therefore the reflection at the right end is very weak.

For the present 1D case as shown in Fig. A1, the total length from $x = 0$ to $x = d$ is divided into *m* parts equally with $\Delta x = d/m$, and the time can be divided into *n* parts equally with $\Delta t = t/n$. By denoting $T_m^n$ as the temperature at the position $x = m\Delta x$ and time $t = n\Delta t$ and using the central difference scheme, Eq. (3) without internal thermal sources can be written as

$$C_{\text{DPL}}^2 \frac{T_{m-1}^n - 2T_m^n + T_{m+1}^n}{(\Delta x)^2} + C_{\text{DPL}}^2 \tau_T \frac{\partial}{\partial t}\left(\frac{T_{m-1}^n - 2T_m^n + T_{m+1}^n}{(\Delta x)^2}\right) = \frac{1}{\tau_q}\frac{\partial T_m^n}{\partial t} + \frac{\partial^2 T_m^n}{\partial t^2}, \quad m=1,2,\ldots M-1. \tag{A10}$$

By using the notations $\partial T_m^n/\partial t = \dot{T}_m^n$ and $\partial^2 T_m^n/\partial t^2 = \ddot{T}_m^n$, Eq. (A10) can be rewritten as

$$(0\ 1\ 0)\begin{pmatrix}\ddot{T}_{m-1}^n\\ \ddot{T}_m^n\\ \ddot{T}_{m+1}^n\end{pmatrix} + \left(-\frac{C_{\text{DPL}}^2 \tau_T}{(\Delta x)^2}\ \ \frac{2C_{\text{DPL}}^2 \tau_T}{(\Delta x)^2} + \frac{1}{\tau_q}\ \ -\frac{C_{\text{DPL}}^2 \tau_T}{(\Delta x)^2}\right)\begin{pmatrix}\dot{T}_{m-1}^n\\ \dot{T}_m^n\\ \dot{T}_{m+1}^n\end{pmatrix} + \left(-\frac{C_{\text{DPL}}^2}{(\Delta x)^2}\ \ \frac{2C_{\text{DPL}}^2}{(\Delta x)^2}\ \ -\frac{C_{\text{DPL}}^2}{(\Delta x)^2}\right)\begin{pmatrix}T_{m-1}^n\\ T_m^n\\ T_{m+1}^n\end{pmatrix} = 0. \tag{A11}$$

If the node $m$ is located at the interface between the sub-layers A and B, the material parameters $\tau_q$, $\tau_T$ and $C_{\text{DPL}}$ are set to be the average values of those for materials A and B. This is a simple approximation of the interface continuity conditions described by Eq. (A4).

At the left boundary, we have

$$(1\ 0)\begin{pmatrix}\ddot{T}_1^n\\ \ddot{T}_2^n\end{pmatrix} + \left(\frac{2C_{\text{DPL}}^2 \tau_T}{(\Delta x)^2} + \frac{1}{\tau_q}\ \ -\frac{C_{\text{DPL}}^2 \tau_T}{(\Delta x)^2}\right)\begin{pmatrix}\dot{T}_1^n\\ \dot{T}_2^n\end{pmatrix} + \left(\frac{2C_{\text{DPL}}^2}{(\Delta x)^2}\ \ -\frac{C_{\text{DPL}}^2}{(\Delta x)^2}\right)\begin{pmatrix}T_1^n\\ T_2^n\end{pmatrix} = \frac{C_n^2}{(\Delta x)^2} T_0^n, \tag{A12}$$

where $T_0^n$ is the temperature at the left boundary (see Eq. (A2)). At the right boundary, we can rewrite the absorbing boundary condition (see Eq. (A9)) as

$$(0\ 1)\begin{pmatrix}\dot{T}_{m-1}^n\\ \dot{T}_m^n\end{pmatrix} + \left(-\frac{C_{\text{DPL}}}{\Delta x}\ \ \frac{C_{\text{DPL}}}{\Delta x}\right)\begin{pmatrix}T_{m-1}^n\\ T_m^n\end{pmatrix} = 0. \tag{A13}$$

Then Eqs. (A11)-(A13) can be combined into a matrix form as

$$\mathbf{E}\begin{bmatrix}\ddot{T}_1^n\\ \vdots\\ \ddot{T}_{m-1}^n\end{bmatrix} + \mathbf{D}\begin{bmatrix}\dot{T}_1^n\\ \vdots\\ \dot{T}_{m-1}^n\end{bmatrix} + \mathbf{C}\begin{bmatrix}T_1^n\\ \vdots\\ T_{m-1}^n\end{bmatrix} = \mathbf{B}. \tag{A14}$$

Applying the first-order backward difference and second-order central difference schemes with respect to time, we can obtain the final equation for the temperature field in the time-domain as

$$\mathbf{E}\frac{\mathbf{T}^{n+1} - 2\mathbf{T}^n + \mathbf{T}^{n-1}}{(\Delta t)^2} + \mathbf{D}\frac{\mathbf{T}^n - \mathbf{T}^{n-1}}{\Delta t} + \mathbf{C}\mathbf{T}^n = 0, \tag{A15}$$

where $\mathbf{T}^n = \begin{bmatrix}T_1^n & \cdots & T_{m-1}^n\end{bmatrix}^T$ and $\mathbf{T}^{n-1} = \begin{bmatrix}T_1^{n-1} & \cdots & T_{m-1}^{n-1}\end{bmatrix}^T$.

problem formulation and verification. Numer Heat Tr B-Fund 2002;41:543.

[40] Kushwaha MS, Halevi P, Martinez G, Dobrzynski L, Djafari-Rouhani B. Theory of acoustic band structure of periodic elastic composites. Phys Rev B 1994;49:2313.

[41] Zhou XZ, Wang YS, Zhang CZ. Effects of material parameters on elastic `s of two-dimensional solid phononic crystals. J App Phys 2009;106:014903.

[42] Krushynska A, Kouznetsova V, Geers M. Visco-elastic effects on wave dispersion in three-phase acoustic metamaterials. J Mech Phys Solids 2016;96:29.

[43] Psarobas I. Viscoelastic response of sonic band-gap materials. Phys Rev B 2001;64:012303.

[44] Engquist B, Majda A. Absorbing boundary conditions for numerical simulation of waves, P Natl Acad Sci USA 1977;74:1765.21